\documentclass[aps,prd,preprint,superscriptaddress,nofootinbib]{revtex4}
\usepackage{amsmath, bm}
\usepackage{graphicx}

\begin{document}

\title{\boldmath 
Resummation of threshold double logarithms \\ 
in gluon fragmentation into $^1S_0^{[1,8]}$ quarkonia}

\author{Hee Sok Chung}
\email{heesokchung@kangwon.ac.kr}
\affiliation{School of Mathematics and Physics, Kangwon National
University, Gangneung 25457, Korea} 

\date{\today}

\begin{abstract}
I investigate the threshold double logarithmic behavior of the 
fragmentation function of a gluon
into $^1S_0^{[1,8]}$ quarkonia, which is suppressed by $1/N$ in
Mellin space. While threshold double logarithms have been resummed to all
orders in the strong coupling for gluon fragmentation functions for quarkonia
at leading power in $1/N$, it has not been known how to resum them for channels
suppressed by powers of $1/N$. Threshold resummation can be important for the
$^1S_0^{[1,8]}$ channels, as the tree-level gluon fragmentation functions
have finite yet nonzero values at the kinematical threshold. 
I derive a formalism for resumming the threshold double logarithms for these
channels at the leading double logarithmic level and discuss the
phenomenological impact. 

\end{abstract}

\maketitle

\section{Introduction}

Quarkonium fragmentation functions (FFs) computed in the nonrelativistic QCD
(NRQCD) factorization formalism~\cite{Bodwin:1994jh} often involve singularities at the kinematical
threshold~\cite{Braaten:1996rp, Braaten:2000pc, Bodwin:2003wh, 
Bodwin:2012xc, Ma:2013yla, Ma:2014eja,
Ma:2015yka, Zhang:2020atv}. The threshold behavior of a FF $D(z)$ can be
quantitatively addressed by considering the Mellin transform
\begin{align}
\tilde{D}(N) = \int_0^1 \frac{dz}{z} z^N D(z), 
\end{align}
where $z$ is the ratio of the momentum of the quarkonium compared to
the fragmenting parton in the $+$ direction. For $D(z)$ to vanish as $z\to1$ 
the Mellin-space FF $\tilde{D}(N)$ must vanish faster than
$1/N$ as $N \to \infty$. If $\tilde{D}(N)$ behaves like $1/N$ at large $N$, 
then $D(z)$ converges to a finite but nonzero value as $z \to 1$. 
If $\tilde{D}(N) \sim 1/N^0$, $D(z)$ is no longer a smooth function in $z$ but
rather a distribution that is singular at $z=1$, such as delta functions and
plus distributions. It is known that the Mellin-space gluon FFs for production
of a quark $Q$ and an antiquark $\bar Q$ in the $^3S_1^{[8]}$, $^3P_J^{[8]}$,
and $^3P_J^{[1]}$ states behave like $1/N^0$ at large $N$, and involve
delta functions and plus distributions singular at $z=1$. 
On the other hand, the gluon FF for the $^3S_1^{[1]}$ state 
vanishes as $z \to 1$~\cite{Braaten:1993rw, Braaten:1995cj, Zhang:2017xoj}. 
In the case of the $^1S_0^{[8]}$ and 
$^1S_0^{[1]}$ channels, the Mellin-space gluon FFs behave like $1/N$ at tree
level, and hence the $z$-space FFs have finite but nonzero values at
$z=1$~\cite{Braaten:1993rw, Braaten:1996rp, Bodwin:2012xc}. 

It has been known that radiative corrections at higher orders in the strong
coupling $\alpha_s$ involve double logarithms in $N$, 
which appear as powers of $\alpha_s \log^2 N$.
These logarithms exacerbate the $z \to 1$ singularities at higher orders in
$\alpha_s$. 
The threshold double logarithms have been resummed for the gluon FFs for the $^3S_1^{[8]}$,
$^3P_J^{[8]}$, and $^3P_J^{[1]}$ channels to all orders in $\alpha_s$ at the
leading double logarithmic accuracy~\cite{Chung:2024jfk, Chung:2026mii}. 
While the gluon FFs computed in fixed-order
perturbation theory involve singular distributions, the resummed FFs vanish
smoothly as $z \to 1$, because the Mellin-space FFs vanish 
faster than any power of $N$ at large $N$. It has been shown that
resummation has a significant impact on the phenomenology of quarkonium 
production at large transverse momentum, especially for differential
observables such as the quarkonium momentum distribution in jets~\cite{Wang:2025drz,
Wang:2026dul, Ha:2026mct}.

For the case of the gluon FFs that are suppressed by $1/N$ or more, 
the threshold logarithms are only known at fixed order for a few channels. 
For example, the corrections at next-to-leading order (NLO) in $\alpha_s$ 
to the gluon FFs for the $^1S_0^{[1,8]}$ channels involve double logarithms in
$1-z$ of the form~\cite{Zhang:2018mlo}
\begin{align}
D_{g \to Q \bar Q (^1S_0^{[1]})} (z) \Big|_{z \to 1} &=
\frac{\alpha_s^2}{2 N_c m^3} \left[ 1 
+ \frac{\alpha_s}{\pi} \left( - \frac{N_c}{2} \log^2(1-z) + \cdots \right)
\right],
\\
D_{g \to Q \bar Q (^1S_0^{[8]})} (z) \Big|_{z \to 1} &=
\frac{(N_c^2-4) \alpha_s^2}{4 N_c m^3} \left[ 1 
+ \frac{\alpha_s}{\pi} \left( - N_c \log^2(1-z) + \cdots \right)
\right].
\end{align}
While the NLO corrections to these FFs have been computed numerically in 
Refs.~\cite{Artoisenet:2014lpa, Artoisenet:2018dbs, Feng:2018ulg}, 
the exact coefficients of the double logarithms have first been reported 
in Ref.~\cite{Zhang:2018mlo}, although such a logarithm 
in the $^1S_0^{[8]}$ channel has already been noted in Ref.~\cite{Artoisenet:2018dbs}. 
Even though the appearance of threshold double logarithms for these channels 
has been established at NLO, it is still not known how to resum the logarithms. 
This is potentially important in heavy
quarkonium production phenomenology, as if the effects of threshold resummation 
are included at leading power in $1/N$ (order $1/N^0$), and not in the 
contributions at order $1/N$, the order-$1/N$ contribution may significantly
affect the threshold behavior of differential observables. This can happen
because while the threshold-resummed gluon FFs for the $^3S_1^{[8]}$,
$^3P_J^{[8]}$, and $^3P_J^{[1]}$ channels vanish quickly as $z \to 1$, 
the gluon FFs for the $^1S_0^{[1,8]}$ channels at tree level has a nonzero
value at $z=1$, and diverge double logarithmically at NLO. 
In practice, inclusion of the effects of 
Dokshitzer-Gribov-Lipatov-Altarelli-Parisi (DGLAP) 
evolution~\cite{Gribov:1972ri, Lipatov:1974qm, Dokshitzer:1977sg, 
Altarelli:1977zs} does make
the Mellin-space FFs for the $^1S_0^{[1,8]}$ channels vanish faster 
than $1/N$ and make the $z$-space FFs vanish at threshold. 
However, because the DGLAP evolution only involves single logarithms, 
resummation of threshold double logarithms may still have a sizable effect 
on the $z$-shapes of the FFs near threshold. 
Thus, in order to bring the quantitative description of quarkonium FFs near
threshold under theoretical control, an all-orders analysis of threshold
logarithms for these channels is necessary.

In this work, I aim to resum the threshold logarithms that appear in the gluon
fragmentation functions for production of $Q \bar Q$ in $^1S_0^{[1,8]}$ states
to all orders in $\alpha_s$, at leading double logarithmic level. 
The method I use is similar to what was used in the previous analysis of
threshold double logarithms for the $^3S_1^{[8]}$,
$^3P_J^{[8]}$, and $^3P_J^{[1]}$ channels in Refs.~\cite{Chung:2024jfk, 
Chung:2026mii}, except that the $^1S_0^{[1,8]}$ channels necessarily involve an
emission of a non-soft gluon in order to produce a spin-singlet state.

This paper is organized as follows. 
I first list the conventions and definitions I use throughout this paper in
Sec.~\ref{sec:conventions}. 
In Sec.~\ref{sec:LOregion}, I compute the 
tree-level fragmentation functions near threshold, which helps identify the
relevant momentum configurations that appear in the FFs. 
In Sec.~\ref{sec:softresum}, 
I identify the origin of the threshold double logarithms 
by considering soft gluon emissions, and obtain a resummation formula at
leading double logarithmic accuracy. 
I present numerical results for the resummed FFs in Sec.~\ref{sec:numer}, 
and conclude in Sec.~\ref{sec:concl}. 

\section{\boldmath Definitions and conventions}
\label{sec:conventions}

Here I collect the definitions and conventions I use throughout this paper. 
I work in dimensional regularization in $d=4-2 \epsilon$ dimensions and use the
Feynman gauge. 
I use the following definition for light-cone coordinates:
\begin{align}
a^\pm = a_\pm = \frac{1}{\sqrt{2}} (a^0 \pm a^3), 
\end{align}
so that $a \cdot b = a^+ b^- + a^- b^+ - \bm{a}_\perp \cdot \bm{b}_\perp$. 
The massless phase space integral for a particle with momentum $k$ is given by 
\begin{align}
\int d{\rm PS}_k &= 
\int_0^\infty \frac{dk^+}{2 \pi} 
\int_0^\infty \frac{dk^-}{2 \pi} 
\int_0^\infty \frac{d^{2-2 \epsilon}k_\perp}{(2 \pi)^{2-2 \epsilon}} 
2 \pi \delta(2 k^+ k^- -\bm{k}_\perp^2)
\nonumber \\
&= \int_0^\infty \frac{dk^+}{4 \pi k^+} 
\int_0^\infty \frac{d^{2-2 \epsilon}k_\perp}{(2 \pi)^{2-2 \epsilon}} 
\end{align}
I define the lightlike vector $n$ through the relation $a^+ = a \cdot n$. 
I adopt the definition for the Collins-Soper FF~\cite{Collins:1981uw} 
used in Ref.~\cite{Chung:2026mii}. 
The adjoint Wilson line in the $k$ direction from spacetime origin to infinity
is defined by 
\begin{align}
\Phi_k = {\cal P} \exp \left[ -i g \int_0^\infty d \lambda \, 
k \cdot A^{\rm adj} (k \lambda) \right]. 
\end{align}
The NRQCD factorization formula for the FF 
$D_{g \to Q \bar Q(^1S_0^{[1,8]})} (z)$ is 
\begin{align}
D_{g \to Q \bar Q(^1S_0^{[1,8]})} (z) &= 
d_{g \to Q \bar Q(^1S_0^{[1,8]})} (z) 
\langle 0 | {\cal O}^{Q \bar Q (^1S_0^{[1,8]})}(^1S_0^{[1,8]}) | 0\rangle, 
\end{align}
where $d_{g \to Q \bar Q(^1S_0^{[1,8]})} (z)$ is the short-distance coefficient
(SDC) and the $\langle 0 | {\cal O}^{Q \bar Q (^1S_0^{[1,8]})}(^1S_0^{[1,8]}) |
0\rangle$ is the NRQCD matrix element. 
The perturbative NRQCD matrix elements are given by~\cite{Braaten:1996rp, 
Bodwin:2012xc, Artoisenet:2018dbs}
\begin{align}
\langle 0 | {\cal O}^{Q \bar Q (^1S_0^{[1]})}(^1S_0^{[1]}) | 0\rangle
&= 2 N_c, \\
\langle 0 | {\cal O}^{Q \bar Q (^1S_0^{[8]})}(^1S_0^{[8]}) | 0\rangle
&= N_c^2-1.
\end{align}
In this work I adopt the normalization for the color-singlet matrix element
as defined in Ref.~\cite{Bodwin:1994jh}, which differs by a factor of 
$2 N_c$ compared to Ref.~\cite{Zhang:2018mlo}. 
The Mellin transform $\tilde{f}$ of a function $f$ is defined as 
\begin{align}
\tilde{f}(N) = \int_0^1 \frac{dz}{z} z^N f(z), 
\end{align}
where all functions of $z$ are defined for $0<z<1$ and vanish for $z>1$. 
The convolution $\otimes$ is defined by 
\begin{align}
(f \otimes g) (z) = \int_z^1 \frac{dz'}{z'} f(z') g(z/z'),
\end{align}
which is diagonalized in Mellin space 
\begin{align}
\int_0^1 \frac{dz}{z} z^N 
(f \otimes g) (z) = \tilde{f}(N) \tilde{g}(N).
\end{align}

\section{\boldmath Tree-level fragmentation functions near threshold}
\label{sec:LOregion}

\begin{figure}[t]
\includegraphics[width=\columnwidth]{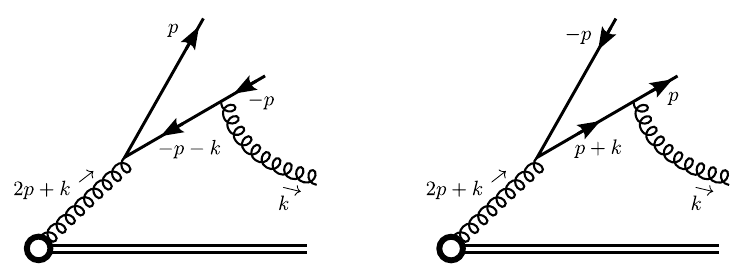}%
\caption{\label{fig:LOdiag}
Amplitude-level diagrams for gluon fragmentation into $Q \bar{Q} g$. 
Solid lines are quarks, curly lines are gluons, 
the open circle denotes the operator that produces a fragmenting gluon in
the Collins-Soper definition of the FF, and the double line represents the 
lightlike Wilson line in the $n$ direction. 
}
\end{figure}

In this section I compute the gluon FFs for production of 
$Q \bar Q (^1S_0^{[1,8]})$ at the lowest nonvanishing order in $\alpha_s$ 
near threshold. 
The production of a $Q \bar Q$ in $^1S_0^{[1,8]}$ states occurs through a
production of a $Q$ and $\bar Q$ from the fragmenting gluon, followed by an
emission of a final-state gluon which is necessary for producing a spin singlet
$Q \bar Q$ in the final state. The amplitude-level diagrams are shown in
Fig.~\ref{fig:LOdiag}.
The subdiagram for the process 
$g^* \to Q \bar Q+g$ is given by 
\begin{align}
{\cal A}_{Q \bar Q g}^{\alpha \mu, a x} &= 
\bar{u}(p) (-i g \gamma^\mu T^x) 
\frac{i (-p\!\!\!/-k\!\!\!/+m)}{(p+k)^2 -m ^2+i \varepsilon}
(-i g \gamma^\alpha T^a) v(p)
\nonumber \\
& \quad + \bar{u}(p) (-i g \gamma^\alpha T^a) 
\frac{i (p\!\!\!/+k\!\!\!/+m)}{(p+k)^2 -m ^2+i \varepsilon}
(-i g \gamma^\mu T^x) v(p), 
\end{align}
where $p=P/2$ is the momentum of the $Q$ or $\bar Q$, $\mu$ and $x$ are the
Lorentz and adjoint indices of the final-state gluon, $k$ is the momentum of
the final-state gluon, $m$ is the heavy quark mass, 
and $\alpha$ and $a$ are the Lorentz and adjoint indices
of the fragmenting gluon. I take nonrelativistic normalization for the
spinors. Since I set the momenta of the $Q$ and $\bar Q$ to be equal, the
only possible orbital angular momentum for the final-state $Q \bar Q$ is
$S$ wave. In order to project onto the spin-singlet state with definite color I
replace the spinors with the following projector~\cite{Barbieri:1975am,
Barbieri:1976fp, Chang:1979nn, Guberina:1980dc, Berger:1980ni}: 
\begin{align}
\bar{u} (p) \Gamma v(p) \to \frac{1}{2 \sqrt{2} m} {\rm tr} [\gamma^5 (p\!\!\!/+m)
\Lambda_c \Gamma], 
\end{align}
where the trace is over color and Dirac matrices, 
$\Gamma$ is any product of Dirac and color matrices, and the color
projector $\Lambda_c$ is given by
\begin{align}
\Lambda_c &= \begin{cases}
\displaystyle \frac{{\bf 1}}{\sqrt{N_c}} & \textrm{color singlet}, \\
\displaystyle \frac{T^c}{\sqrt{T_F}} & \textrm{color octet},
\end{cases}
\end{align}
with $N_c = 3$ the number of colors, $\bf 1$ the $SU(N_c)$ unit matrix, 
$c$ is the adjoint index carried by the color-octet state, 
and $T_F = 1/2$. 
The trace over Dirac and color matrices yields 
\begin{align}
{\cal A}_{Q \bar Q(^1S_0^{[1]}) g}^{\alpha \mu, a x} &=
\frac{\sqrt{2} g^2}{m} 
\frac{\epsilon^{\mu \nu \alpha \beta} p_\beta k_\nu }{2 k \cdot p+i
\varepsilon} \frac{2 T_F \delta^{ax} }{\sqrt{N_c}}, \\
{\cal A}_{Q \bar Q(^1S_0^{[8]}) g}^{\alpha \mu, a c x} &=
\frac{\sqrt{2} g^2}{m} 
\frac{\epsilon^{\mu \nu \alpha \beta} p_\beta k_\nu }{2 k \cdot p+i
\varepsilon} d^{acx} \sqrt{T_F},
\end{align}
where I set $p^2= m^2$ and $k^2 =0$. 

It is worth examining the dependence of the subdiagram on the 
final-state gluon momentum $k$. 
Due to the Levi-Civita tensor, $k$ must be orthogonal to the polarizations of
both the final-state and the fragmenting gluons; further, if I take the frame
where $\bm{p}_\perp = \bm{0}$, then $\bm{k}_\perp$ cannot appear in the
numerator. As a result, the numerator only involves $k^-$ 
in the threshold region since $k^+ \to 0$. 
Then the amplitude becomes proportional to 
\begin{align}
\frac{k^-}{2 k \cdot p + i \varepsilon} 
\approx 
\frac{k^-}{2 k^- p^+ + i \varepsilon},
\end{align}
where I neglected $k^+ p^-$ compared to $k^- p^+$ in the threshold region.
Since the $k^-$ cancels between the numerator and the denominator, the
subdiagram ${\cal A}_{Q \bar Q(^1S_0^{[1, 8]}) g}$ cannot supply any scale for
the phase space integral over $k$. Moreover, because the gluon with momentum
$k$ carries a transverse polarization, the soft approximation does
not apply to this gluon. 

The SDC for the tree-level gluon FF is obtained by squaring the subdiagram 
${\cal A}_{Q \bar Q(^1S_0^{[1, 8]}) g}$, 
multiplying by the Feynman rules for the fragmenting gluon, 
summing over the colors and polarizations of the initial and final states, 
and finally dividing by the normalization for the NRQCD LDME in perturbation
theory. I obtain, in the threshold region, 
\begin{align}
d_{g \to Q \bar Q(^1S_0^{[1,8]})}^{\rm threshold} (z) &= 
{\cal C}^{[1,8]}
\frac{16 \pi \alpha_s^2}{m} 
\int d{\rm PS}_k 
\frac{P^+ 2 \pi \delta(k^+ - P^+ (1-z))}{|(k+P)^2+i \varepsilon|^2},
\end{align}
where ${\cal C}^{[1]} = 1/(2 N_c^2)$ and ${\cal C}^{[8]} = (N_c^2-4)/[2 N_c
(N_c^2-1)]$. The normalization for the color-singlet channel differs from
Ref.~\cite{Zhang:2018mlo} by a factor of $2 N_c$ due to a different choice of
the normalization of the color-singlet matrix element; 
in this work I adopt the definition in Ref.~\cite{Bodwin:1994jh}. 
The phase-space integral over $k$ is computed in the threshold region as 
\begin{align}
\label{eq:kphase}
\int d{\rm PS}_k
\frac{P^+ 2 \pi \delta(k^+ - P^+ (1-z))}{|(k+P)^2+i \varepsilon|^2}
&\approx
\int_0^\infty \frac{dk^+ }{4 \pi k^+ } 
\int \frac{d^{2-2 \epsilon} k_\perp}{(2 \pi)^{2-2 \epsilon}} 
\frac{P^+ 2 \pi \delta(k^+ - P^+ (1-z))}
{\left( \bm{k}_\perp^2 P^+/k^+ + P^2 \right)^2}
\nonumber \\ &= 
\frac{(1-z)^{-\epsilon} \Gamma(1+\epsilon)}{32 \pi m^{2}} 
\left( \frac{\pi \mu^2}{m^2} \right)^{\epsilon} ,
\end{align}
from which I obtain 
\begin{align}
\label{eq:DLO}
d_{g \to Q \bar Q(^1S_0^{[1,8]})}^{\rm threshold} (z) &=
{\cal C}^{[1,8]}
\frac{\alpha_s^2}{2 m^3}
(1-z)^{-\epsilon} \Gamma(1+\epsilon)
\left( \frac{\pi \mu^2}{m^2} \right)^{\epsilon} .
\end{align}
This result agrees with Refs.~\cite{Braaten:1996rp, Bodwin:2012xc} 
for $d=4$ in the $z \to 1$ limit. 
In the case of arbitrary spacetime dimensions, the results  in
Refs.~\cite{Artoisenet:2014lpa, Zhang:2018mlo} 
contain an additional factor of $d-3= 1-2 \epsilon$ which becomes
unity as $\epsilon \to 0$. This happens because 
I only kept the $\epsilon$ dependences originating from the phase
space integral while the Dirac trace has been computed strictly in 4 spacetime
dimensions. The $d$-dimensional calculation of the Dirac trace requires
choosing a definition of the spin-singlet state in arbitrary spacetime
dimensions; the exact factor $1-2 \epsilon$ is obtained by adopting the 
definition used in Ref.~\cite{Artoisenet:2014lpa} when computing the Dirac
trace involving $\gamma^5$ in $d$ dimensions. Since this additional
factor depends only on the $d$-dimensional definition of the final-state 
$Q \bar Q$, it cannot depend on $z$ and does not affect the threshold double 
logarithms I consider in this work.

\section{\boldmath Threshold double logarithms}
\label{sec:softresum}

Now I consider the FFs at NLO in $\alpha_s$ and investigate the origin of the
threshold double logarithms in order to obtain a resummation formalism for the 
gluon FFs for the $^1S_0^{[1,8]}$ channels. 
This is carried out by expanding loop corrections in powers of $1-z$. 
Since the goal of this work is to resum double logarithms, I only retain the 
leading contributions in the expansion and discard contributions that are 
suppressed by powers of $1-z$. 
Compared to the analyses in Refs.~\cite{Chung:2024jfk, Chung:2026mii}, 
the $^1S_0^{[1]}$ and $^1S_0^{[8]}$ FFs involve one more complication due to
the appearance of the final-state gluon with momentum $k$ from tree level. 
In the $Q \bar Q$ rest frame, 
the typical scaling of momentum $k$ is such that $k^+ \sim M(1-z)$, while 
$\bm{k}_\perp^2 \sim M^2 (1-z)$, and $k^- \sim M$. 
That is, $k$ is collinear to minus, similarly to the lightlike Wilson line in
the $n$ direction. However, we will see later that the two directions must be
kept distinct, as logarithms arise from gluon exchanges between the two
lightlike directions. 

\subsection{Organization of soft gluons}

I first consider emission of soft gluons from the outermost quark and antiquark
lines. I apply the Grammer-Yennie approximation~\cite{Grammer:1973db} as
explained in Refs.~\cite{Chung:2024jfk, Chung:2026mii}. 
In the color-singlet case, the effects of soft gluon emissions cancel between
the quark and antiquark lines; in the color-octet case, the soft-gluon
attachments can be replaced by attachments to an adjoint Wilson line in the $p$
direction. 

Then I consider emission of soft gluons from the final-state gluon with
momentum $k$. The Feynman rule for the three-gluon vertex is given by 
\begin{align}
V_{\mu_{\rm in} \mu_{\rm out} \mu_s}^{a_{\rm in} a_{\rm out} a_s} (k+l,-k,-l)
&= g f^{a_{\rm in} a_s a_{\rm out}}
[ g^{\mu_{\rm in} \mu_s} (k+2 l)^{\mu_{\rm out}}
+ g^{\mu_{\rm out} \mu_s} (k-l)^{\mu_{\rm in}}
\nonumber \\ & \hspace{14ex} 
- g^{\mu_{\rm in} \mu_{\rm out}} (2 k+l)^{\mu_s} ]
\end{align}
where the gluon with indices $\mu_{\rm in}$, $a_{\rm in}$ flows into the vertex
with momentum $k+l$, and the gluon with momentum $k$ flows out from the vertex 
carrying indices $\mu_{\rm out}$, $a_{\rm out}$; the gluon with momentum $l$
flows out from the vertex carrying indices $\mu_s$, $a_s$. 
When $l$ is soft compared to $k$, this vertex can be approximated by 
\begin{align}
V_{\mu_{\rm in} \mu_{\rm out} \mu_s}^{a_{\rm in} a_{\rm out} a_s} (k+l,-k,-l)
&\approx g f^{a_{\rm in} a_s a_{\rm out}}
\left( g^{\mu_{\rm in} \mu_s} k^{\mu_{\rm out}}
+ g^{\mu_{\rm out} \mu_s} k^{\mu_{\rm in}}
- 2 g^{\mu_{\rm in} \mu_{\rm out}} k^{\mu_s} \right). 
\end{align}
The first two terms in the parenthesis do not contribute to the FF because the
momentum $k$ from the vertex becomes dotted with the Lorentz index of the gluon
with momentum $k$. When combined with the propagator of the gluon with momentum
$k+l$, which is given by $-i/(2 k\cdot l+i \varepsilon)$, the contribution from
the last term in the parenthesis is equivalent to the gluon attachment to an
adjoint Wilson line in the $k$ direction. 

In general, soft gluons may also attach onto the virtual quark line with
momentum $\pm (p+k)$, as well as the fragmenting gluon with momentum $P+k$. 
However, since these involve direct attachments of soft lines onto a
hard subdiagram, the contributions from these types of soft gluon attachments
are suppressed by powers of $1-z$ and can be discarded. 

There is one configuration of diagrams that cannot be obtained by adding gluon
exchanges to the tree-level diagram. This involves the $Q \bar Q g$ being 
produced from the emission of a gluon from the $n$-direction Wilson line. 
In this case, the fragmenting gluon must carry a momentum that is either
collinear or soft. In both cases, these kinds of diagrams involve additional
attachments of soft or collinear lines onto a hard subdiagram, and these
contributions are power suppressed.

The effect of soft-gluon emissions at leading power in $1-z$ can then be
represented by adjoint Wilson lines. 
In the color-singlet case, the soft contributions can be represented by the
following operator at the amplitude level:
\begin{align}
\frac{\delta^{xa}}{\sqrt{N_c^2-1}} T \left( \Phi_k^{xx'} \Phi_n^{aa'} \right), 
\end{align}
where $T$ denotes time ordering. 
In the color-octet case, the operator is 
\begin{align}
\frac{\sqrt{N_c} d^{acx} }{\sqrt{(N_c^2-1) (N_c^2-4)}}
T \left( \Phi_p^{cc'} \Phi_k^{xx'} \Phi_n^{aa'} \right).
\end{align}
The color factors are chosen so that $(\delta^{xa}/\sqrt{N_c^2-1})^2 = 
\left( \sqrt{N_c} d^{acx}/\sqrt{(N_c^2-1) (N_c^2-4)} \right)^2=1$. 
It is tempting to proceed to define soft functions by the vacuum expectation 
values of 
squares of these operators. However, as will become apparent in the next
section, when integrating over the phase space the momentum of the gluon
exchanged between the Wilson lines over the final-state cut does not decouple
trivially from the fragmenting gluon propagator denominator, which prohibits
the simple form of factorization that was obtained for the FFs in 
Refs.~\cite{Chung:2024jfk, Chung:2026mii}. 
In the next section, I will discuss how the soft functions appear in the
FFs in the threshold region by considering diagrams at one-loop level.

\subsection{Threshold double logarithms at NLO}

Now I compute diagrams that give rise to threshold double logarithms at NLO. 
For double logarithms to appear, the NLO calculation must involve a double pole
in $\epsilon$. In calculation of scattering amplitudes, double poles
appear from the overlap of soft and collinear divergences. Hence, threshold
double logarithms must arise from gluon exchanges between two lines of distinct
directions, with at least one being lightlike. The simplest diagram where this
occurs comes from the exchange of a gluon between Wilson lines in the $n$ and
$p$ directions, which only contributes to the color-octet channel. 
The contribution from this diagram to $d_{g \to Q \bar Q(^1S_0^{[8]})}$ is
given by 
\begin{align}
{\cal C}^{[8]} \frac{16 \pi \alpha_s^2}{m} \times 2 \, {\rm Re} I_{pn}, 
\end{align}
where\footnote{Here I do not distinguish between the ultraviolet and infrared
poles, because such a distinction is not necessary in obtaining the finite 
pieces at order $\epsilon^0$. This way, the virtual diagram is scaleless and
can be discarded.} 
\begin{align}
I_{pn} &=
g^2 C_A 
\int d{\rm PS}_k
\int d{\rm PS}_l
\frac{P^+ 2 \pi \delta(k^+ +l^+ - P^+ (1-z))}
{ [(k+l+P)^2+i \varepsilon] [(k+P)^2-i \varepsilon] }
\frac{P \cdot n}{(P \cdot l+i \varepsilon) (l \cdot n-i \varepsilon)}.
\end{align}
Here the factor $C_A = N_c = 3$ comes from the 
Casimir in the adjoint representation 
$f^{acd} f^{bcd} = C_A \delta^{ab}$. 
Note that if the gluon with momentum $l$ is attached to the $p$-direction
Wilson line, the momentum of the fragmenting gluon is $k+l+P$. On the other
hand, if the gluon with momentum $l$ is attached to the $n$-direction Wilson
line, the momentum of the fragmenting gluon is instead $k+P$.
It is straightforward to compute the integral $I_{pn}$ 
in the threshold region by using the technique explained in Appendix C of
Ref.~\cite{Zhang:2018mlo},
which is based on the method of regions. 
It is most convenient to work in a frame where $\bm{P}_\perp = \bm{0}$. 
Following Ref.~\cite{Zhang:2018mlo}, I rewrite the massless two-particle 
phase space integral by integrating out the $2 \pi \delta(k^++l^+-P^+(1-z))$ to
eliminate $l^+$ and change the integration variable from $k^+$ to $z_1$ through
the rescaling $z_1 = k^+/[(1-z) P^+]$. 
I obtain
\begin{align}
& \hspace{-5ex}
\int d{\rm PS}_k \int d{\rm PS}_l \, P^+ 2 \pi \delta(k^+ +l^+ - P^+ (1-z))
\nonumber \\ 
&= 
\frac{1}{8 \pi (1-z)}
\int_0^1 \frac{dz_1}{z_1 (1-z_1)} 
\int \frac{d^{d-2} k_\perp}{(2 \pi)^{d-2}}
\int \frac{d^{d-2} l_\perp}{(2 \pi)^{d-2}} 
\equiv \int d \Phi_2,
\end{align}
where $k^- = \bm{k}_\perp^2/(2 k^+)$, $l^- = \bm{l}_\perp^2/(2 l^+)$, 
$k^+ = (1-z) z_1 P^+$, and $l^+ = (1-z) (1-z_1) P^+$. 
The scale of the $k_\perp$ and the $l_\perp$ integrals are determined by the
denominator factors. For example, the $(k+P)^2$ denominator gives 
\begin{align}
(k+P)^2 
&= \frac{1}{(1-z)} 
\left[ \frac{\bm{k}_\perp^2}{z_1} + (1-z) [1+ (1-z) z_1] M^2 \right],
\end{align}
where $M^2 = 2 P^+ P^- = 4 m^2$. 
The $(k+l+P)^2$ denominator gives 
\begin{align}
(k+l+P)^2
&= \frac{1}{(1-z)}
\left[ \frac{\bm{k}_\perp^2}{z_1} + 
\frac{\bm{l}_\perp^2}{1-z_1} + (1-z)[1+ (1-z) ] M^2 \right]
\nonumber \\
& \quad + 
\left( \sqrt{\frac{z_1}{1-z_1}} \bm{l}_\perp - \sqrt{ \frac{1-z_1}{z_1}}
\bm{k}_\perp \right)^2, 
\end{align}
and the $P\cdot l$ denominator gives 
\begin{align}
P \cdot l 
&= \frac{1}{2(1-z)}
\left[ \frac{\bm{l}_\perp^2}{1-z_1} + (1-z)^2 (1-z_1) M^2 \right]. 
\end{align}
It is apparent that while the only relevant region for the $\bm{k}_\perp$ 
integral is $\bm{k}_\perp^2 \sim M^2 (1-z)$, the $\bm{l}_\perp$ integral 
involves two regions $\bm{l}_\perp^2 \sim M^2 (1-z)$ and 
$\bm{l}_\perp^2 \sim M^2 (1-z)^2$.
The threshold behavior of $I_{pn}$ is given by the sum of the contributions
from the two regions:
\begin{align}
I_{pn}^{\rm threshold} 
&=
g^2 C_A \, \int d \Phi_2 \frac{2 (1-z)^2}{ (1-z_1)
\left[ \frac{\bm{k}_\perp^2}{z_1} + (1-z) M^2 \right]
\left[ \frac{\bm{k}_\perp^2}{z_1} +
\frac{\bm{l}_\perp^2}{1-z_1} + (1-z) M^2 \right]
\left[ \frac{\bm{l}_\perp^2}{1-z_1} \right] }
\nonumber \\ &  \quad + 
g^2 C_A \, \int d \Phi_2 \frac{2 (1-z)^2}{ (1-z_1)
\left[ \frac{\bm{k}_\perp^2}{z_1} + (1-z) M^2 \right]^2
\left[ \frac{\bm{l}_\perp^2}{1-z_1} + (1-z)^2 (1-z_1) M^2 \right] }, 
\end{align}
where in the first integral $M^2 (1-z)^2$ is neglected compared to
$\bm{l}_\perp^2$, and in the second integral $\bm{l}_\perp^2 \sim M^2 (1-z)^2$ 
is neglected compare to $\bm{k}_\perp^2$. 
The result is 
\begin{align}
I_{pn}^{\rm threshold}
&=
\frac{(\mu/M)^{4 \epsilon}}{32 \pi m^2} 
\frac{\alpha_s C_A }{\pi }
\left[ \frac{(1-z)^{-2 \epsilon}}{\epsilon^2} 
- \frac{(1-z)^{-3 \epsilon}}{2 \epsilon^2} + \cdots \right]
\nonumber \\
&= 
\frac{(\mu/M)^{4 \epsilon}}{32 \pi m^2} 
\frac{\alpha_s C_A }{\pi }
\left[ \frac{1}{2 \epsilon^2} - \frac{\log (1-z)}{2 \epsilon}
- \frac{\log^2 (1-z)}{4} + O(\epsilon) \right]
\end{align}
where I neglect all higher order terms in $\epsilon$ that do not involve
$1-z$, $m$, or $\mu$ other than the double pole that generates the threshold 
double logarithm. In the last line, I expanded in powers of $\epsilon$ to
unveil the $\log^2(1-z)$ term. 

It is worth carefully examining the results from the two regions. 
Since $l^+ \sim (1-z) M$, $\bm{l}_\perp^2 \sim (1-z) M^2$ implies that 
$l^- \sim M$, meaning that the region $\bm{l}_\perp^2 \sim (1-z) M^2$ actually
corresponds to the collinear region. The soft region comes from $\bm{l}_\perp^2
\sim (1-z)^2 M^2$, so that $l^+ \sim l^- \sim (1-z) M$. 
That is, the complete contribution from one-gluon exchange between $\Phi_p$ and
$\Phi_n$ are the sum of soft and collinear contributions. 
Although the $p$-direction Wilson line was obtained by organizing soft gluon
attachments to the quark and antiquark lines, the integral $I_{pn}$ itself is
valid in both soft and collinear regions. Nevertheless, the collinear region
cannot be discarded. 

The collinear region is problematic for two reasons. Since in this region 
$l$ cannot be neglected compared to $k$ in the denominators, 
the $k$ and $l$ integrals do not decouple. Also, for the same reason, the
approximation that was used in the previous section to organize 
soft gluon emissions 
into the Wilson line $\Phi_k$ fails in the collinear region.
The latter issue is critical, as the extra terms that survive in the collinear
region cannot be generalized to arbitrarily many gluon exchanges in a
straightforward manner, unlike the soft region. 

Fortunately, 
since the only goal of this work is to reproduce the threshold double
logarithmic term at order $\epsilon^0$, there is a simple workaround. 
Note that because the scaling of the
collinear and soft regions are fixed by the $\bm{k}_\perp$ and $\bm{l}_\perp$ 
integrals, all of the logarithms in $1-z$ will disappear if I set 
$\mu^2 = M^2 (1-z)$ for the collinear region and 
$\mu^2 = M^2 (1-z)^2$ for the soft region. 
The $\log^2 (1-z)$ term is then restored by evolving the scales to a common
value. At fixed order this is most easily accomplished by setting 
$\mu^2 = M^2 (1-z)$ for both collinear and soft regions, which also 
automatically discards the logarithmic contributions from loop corrections in
the collinear region. 
As a result, the $\log^2(1-z)$ term can be obtained simply by considering loop
integration only in the soft region, then setting all scales to $\mu^2= M^2
(1-z)$. That is, 
\begin{align}
I_{pn}^{\rm soft} &=
g^2 C_A
\int_0^\infty \frac{d \xi }{2 \pi} 
\int d{\rm PS}_k
\frac{P^+ 2 \pi \delta(k^+ +\xi - P^+ (1-z))} { |(k+P)^2+i \varepsilon|^2 }
\nonumber \\ & \hspace{20ex} \times 
\int d{\rm PS}_l
\frac{2 \pi \delta (l^+ - \xi) 
P \cdot n}{(P \cdot l+i \varepsilon) (l \cdot n-i \varepsilon)}.
\end{align}
Here I split the delta function into a convolution integral of two delta 
functions in order to rewrite the integral in a factorized form. 
The $l$ integral can be regarded as a one-loop contribution from the soft
function coming from the exchange of a gluon between $\Phi_p$ and $\Phi_n$,
while the $k$ integral is the tree-level hard-collinear contribution. 
This evaluates to 
\begin{align}
I_{pn}^{\rm soft, \,threshold}|_{\mu^2 = M^2 (1-z)}
&= 
\frac{(1-z)^{2 \epsilon}}{32 \pi m^2}
\frac{\alpha_s C_A }{\pi }
\left[ 
- \frac{(1-z)^{-3 \epsilon}}{2 \epsilon^2} + \cdots \right]
\nonumber \\
&= 
\frac{1}{32 \pi m^2}
\frac{\alpha_s C_A }{\pi }
\left[ 
-\frac{1}{2 \epsilon^2} + \frac{\log(1-z)}{2 \epsilon} - \frac{\log^2(1-z)}{4}
+O(\epsilon)
\right], 
\end{align}
which exactly reproduces the threshold double logarithm in the full result.

The double pole coming from the soft $l$ integral can be understood in terms of
the cusp anomalous dimension at one loop. In general, a cusp between two 
non-lightlike directions $p_1$ and $p_2$ produces a pole proportional to 
$\epsilon^{-1} \log \frac{(2 p_1 \cdot p_2)^2}{p_1^2 p_2^2}$ for large cusp
angles~\cite{Korchemsky:1987wg}. 
When one of the directions is taken to be exactly lightlike, the collinear
divergence produces another pole in $\epsilon$, thus giving a double
pole~\cite{Korchemskaya:1992je, Bassetto:1993xd, Korchemsky:1988si,
Korchemsky:1993uz, Drummond:2007aua, Cherednikov:2012yd}. 
The dependence on the directions of $p_1$ and $p_2$ appear as an overall factor
of $(p_1 \cdot p_2)^\epsilon$, which is fixed by Lorentz invariance and the
scaling behavior with respect to the the lightlike momentum, so that it only
affects higher order terms in $\epsilon$. 

It is worth emphasizing that this does not mean that the loop corrections in 
the collinear region vanish with the scale choice $\mu^2 = M^2 (1-z)$. 
However, because the only goal is to reproduce the double logarithms in $1-z$, 
and setting $\mu^2 = M^2 (1-z)$ removes all double logarithms that arise from
the collinear region, I can discard the loop corrections from this region
without affecting the double logarithms with this specific scale choice. 

Note that this procedure described above can be applied for a general loop 
diagram that
gives rise to a double logarithm in $1-z$ at the leading double logarithmic
level. As argued previously, the leading double logarithms can only arise from
gluon exchanges between lines of two distinct directions, with at least one
being lightlike. For multiple gluon exchanges, only the planar diagrams give
rise to the leading double logarithms, because only for the planar
configurations the gluon and Wilson-line propagators have leading power counts
in the expansion in $1-z$. 
For these diagrams, the method of
regions can be applied to express the leading contribution in the threshold
region in terms of collinear and soft loop integrals. The scale of
the collinear loop integral is $\mu^2 = M^2 (1-z)$, so that a collinear 
$n$-loop integral can produce leading double logarithms of the form 
$\alpha_s^n [M^2(1-z)/\mu^2]^{-n \epsilon} \epsilon^{-2 n}$. Hence, all double
logarithms from the collinear region disappear for the scale choice 
$\mu^2 = M^2 (1-z)$.

\begin{figure}[t]
\includegraphics[width=\columnwidth]{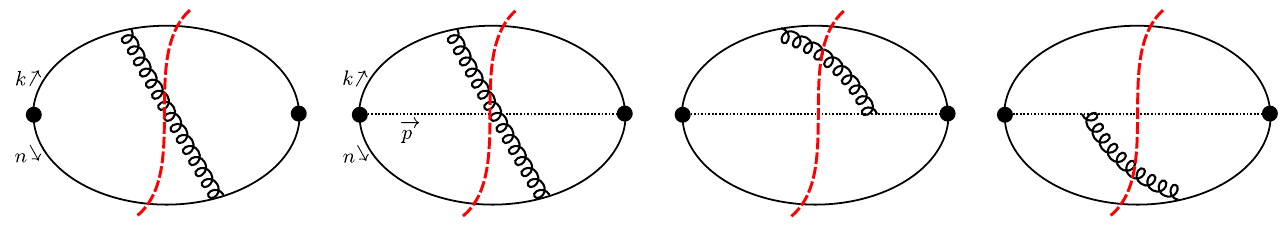}%
\caption{\label{fig:softNLO}
Graphical representations of the soft functions at one loop. 
Solid lines are lightlike Wilson lines, dotted lines are temporal Wilson lines,
curly lines are gluons, and the dashed lines represent the final-state
cut. The arrows denote the directions of the Wilson lines. The first diagram is
the one-loop correction to the $^1S_0^{[1]}$ soft function, while the remaining
diagrams come from the $^1S_0^{[8]}$ soft function. 
}
\end{figure}

The result for $I_{pn}^{\rm soft}$ can already be generalized to write down the
FFs in the threshold region at the leading double logarithmic accuracy:
\begin{align}
d_{g \to Q \bar{Q} (^1S_0^{[1,8]})}^{\rm DL} (z) 
&= {\cal C}^{[1,8]} \frac{16 \pi \alpha_s^2}{m} 
\int \frac{d \xi}{2 \pi} 
J(P^+ (1-z)-\xi,\mu) S^{[1,8]} (\xi,\mu) \Big|_{\mu^2 = M^2 (1-z)}, 
\end{align}
where $J(\xi,\mu)$ is the hard-collinear contribution given at leading order
(LO) in $\alpha_s$ by 
\begin{align}
J(\xi,\mu) = \int d{\rm PS}_k 
\frac{P^+ 2 \pi \delta(k^+ -\xi)} { |(k+P)^2+i \varepsilon|^2 }, 
\end{align}
and the $S^{[1,8]} (\xi,\mu)$ are the soft functions 
\begin{align}
S^{[1]} (\xi,\mu) &= 
\frac{1}{N_c^2-1} 
\langle 0 | \bar{T} 
\left( \delta^{yb} \Phi_k^{yx'} \Phi_n^{ba'} \right) ^\dag
2 \pi \delta(\hat{p} \cdot n - \xi) 
T \left( \delta^{xa} \Phi_k^{xx'} \Phi_n^{aa'} \right) | 0 \rangle, 
\\
S^{[8]} (\xi,\mu) &=
\frac{N_c}{(N_c^2-1) (N_c^2-4)}
\langle 0 | \bar{T}
\left( 
d^{bdy} \Phi_p^{dc'} \Phi_k^{yx'} \Phi_n^{ba'}
\right) ^\dag 
2 \pi \delta(\hat{p} \cdot n - \xi)
\nonumber \\ & \hspace{25ex} \times 
T \left( 
d^{acx} \Phi_p^{cc'} \Phi_k^{xx'} \Phi_n^{aa'}
\right) | 0 \rangle, 
\end{align}
where $\bar{T}$ denotes antitime ordering, and $\hat{p}$ is an operator that
reads off the momentum of the operator to the right. 
Because of momentum conservation, the gluons in the soft
region are always closer to the final state than the gluons in the collinear
region, so that the leading double logarithmic loop corrections to the 
soft region can be computed from the above soft functions. 
Although the soft functions depend on the direction $k$ of the Wilson line
$\Phi_k$, as mentioned beforehand the double poles that produce the threshold
double logarithms do not depend on the Wilson-line directions, so that the $k$
integral decouples in the leading double logarithmic level. 
To NLO accuracy, the results are 
\begin{align}
d_{g \to Q \bar{Q} (^1S_0^{[1]})}^{\rm DL} (z)
&= 
{\cal C}^{[1]} \frac{\alpha_s^2}{2 m^3} 
\left[ 1 - \frac{\alpha_s C_A}{\pi} 
(1-z)^{-\epsilon} 
\frac{1}{\epsilon^2} + O(\alpha_s^2)
\right], 
\\
d_{g \to Q \bar{Q} (^1S_0^{[8]})}^{\rm DL} (z)
&=
{\cal C}^{[8]} \frac{\alpha_s^2}{2 m^3}
\left[ 1 - \frac{\alpha_s C_A}{\pi} 
(1-z)^{-\epsilon}
\left( 
\frac{1}{\epsilon^2} 
+
\frac{1}{2 \epsilon^2} 
+
\frac{1}{2 \epsilon^2} 
\right) 
+ O(\alpha_s^2)
\right],
\end{align}
where the double pole in the color-singlet channel comes from the gluon
exchange between $\Phi_k$ and $\Phi_n$, while the double poles in the
color-octet channel come from gluon exchanges between 
$\Phi_k$ and $\Phi_n$, $\Phi_p$ and $\Phi_n$, and $\Phi_p$ and $\Phi_k$, 
respectively. 
Graphical representations of the soft functions at one loop are shown in
Fig.~\ref{fig:softNLO}. 
The double pole from the gluon exchange between $\Phi_k$ and $\Phi_n$ has a
coefficient that is twice as large as other exchanges, because in this case 
both the $k$ and $n$ directions are lightlike and produce collinear poles. 
It is not difficult to compute the contribution from the gluon exchange between
$\Phi_p$ and $\Phi_k$ directly in momentum space, which results in the same
integral as $I_{pn}$ in the threshold region. For the case of the gluon
exchange between $\Phi_k$ and $\Phi_n$, a momentum-space calculation can be
quite involved due to the rapidity divergence that requires regularization. 
Instead, the double pole itself can be obtained straightforwardly by using a
position-space representation of the gluon propagator~\cite{Beneke:2009rj, 
Laenen:2015jia}. 

The double logarithmic term at NLO can be found by expanding in powers of
$\epsilon$ and keeping only the $\log^2 (1-z)$ term at NLO. The result is 
\begin{align}
\label{eq:thresholdNLOsinglet}
d_{g \to Q \bar{Q} (^1S_0^{[1]})}^{\rm DL, \, finite} (z)
&=
{\cal C}^{[1]} \frac{\alpha_s^2}{2 m^3}
\left[ 1 - \frac{\alpha_s C_A}{\pi} \frac{\log^2(1-z)}{2} + O(\alpha_s^2)
\right],
\\
\label{eq:thresholdNLOoctet}
d_{g \to Q \bar{Q} (^1S_0^{[8]})}^{\rm DL, \, finite} (z)
&=
{\cal C}^{[8]} \frac{\alpha_s^2}{2 m^3}
\left[ 1 - \frac{\alpha_s C_A}{\pi} \log^2(1-z)
+ O(\alpha_s^2)
\right].
\end{align}
These agree completely with the results found in the fixed-order 
calculation in Ref.~\cite{Zhang:2018mlo}.

\subsection{Resummation}

At higher orders in $\alpha_s$ the results in 
Eqs.~(\ref{eq:thresholdNLOsinglet}) and (\ref{eq:thresholdNLOoctet}) 
generalize to 
planar diagrams involving gluon exchanges between the Wilson lines, 
which can be immediately exponentiated in Mellin space. 
The form of the Mellin-space exponent commonly used in threshold resummation is 
\begin{align}
\int_0^1 \frac{dz}{z} z^N \left[ \frac{-2 \log(1-z)}{1-z} \right]_+
= - \psi^{(0)} (N) \left( \psi^{(0)} (N) + 2 \gamma_{\rm E} \right)
+ \psi^{(1)} (N) - \frac{\pi^2}{6} - \gamma_{\rm E}^2, 
\end{align}
where $\psi^{(0)}(z) = \Gamma'(z)/\Gamma(z)$ is the digamma function, 
$\psi^{(1)} (z) = d \psi^{(0)}(z)/dz$, 
and $\gamma_{\rm E}$ is the Euler–Mascheroni constant. 
For large $N$, this form behaves like $-\log^2 N$. 
This form has the advantage that it preserves the total fragmentation
probability, as it vanishes at $N=1$. 
To obtain the resummed expression I first replace the NLO 
double logarithmic term $-\log^2(1-z)$ with the expression 
\begin{align}
\left[ \frac{-2 \log(1-z)}{1-z} \right]_+ \otimes 1
= - \log^2 (1-z) + 2\, {\rm Li}_2 (1-z) + 2 \log z \log(1-z),
\end{align}
where ${\rm Li}_2 (z) = \sum_{k=1}^\infty z^k/k^2$ is the dilogarithm. 
For $z \to 1$ the extra terms vanish
as $2\, {\rm Li}_2 (1-z) + 2 \log z \log(1-z) = 
2 (1-z) [1-\log (1-z)] + O((1-z)^2)$, so that this replacement does not modify
the logarithms at leading power in $1-z$. 
The resummed fragmentation functions are then given by 
\begin{align}
\label{eq:resum}
\tilde{d}_{g \to Q \bar{Q} (^1S_0^{[1,8]})}^{\rm Resum} (N)
&=
\exp \left[ J^N_{^1S_0^{[1,8]}} \right] 
\tilde{d}_{g \to Q \bar{Q} (^1S_0^{[1,8]})}^{\rm LO} (N), 
\end{align}
with 
\begin{align}
J^N_{^1S_0^{[8]}} = 
\frac{\alpha_s C_A}{\pi} 
\int_0^1 \frac{dz}{z} z^N \left[ \frac{-2 \log(1-z)}{1-z} \right]_+, 
\quad 
J^N_{^1S_0^{[1]}} = \frac{1}{2} J^N_{^1S_0^{[8]}}. 
\end{align}
If I combine the resummed result with the NLO correction from the
fixed-order calculation, I obtain the expression
\begin{align}
\label{eq:NLOresum}
\tilde{d}_{g \to Q \bar{Q} (^1S_0^{[1,8]})}^{\rm NLO+ Resum} (N)
&=
\exp \left[ J^N_{^1S_0^{[1,8]}} \right]
\left( \tilde{d}_{g \to Q \bar{Q} (^1S_0^{[1,8]})}^{\rm NLO} (N)
- J^N_{^1S_0^{[1,8]}}  \tilde{d}_{g \to Q \bar{Q} (^1S_0^{[1,8]})}^{\rm LO} (N)
\right). 
\end{align}
where $\tilde{d}_{g \to Q \bar{Q} (^1S_0^{[1,8]})}^{\rm NLO} (N)$ is the 
gluon FF computed at fixed-order perturbation theory to NLO accuracy, which
includes both the LO and NLO contributions. 
The term $J^N_{^1S_0^{[1,8]}} \tilde{d}_{g \to Q \bar{Q}
(^1S_0^{[1,8]})}^{\rm LO} (N)$ subtracts the double logarithmic contribution in
the fixed-order NLO calculation so that there is no double counting. 

The forms of the resummed FFs I obtain in Eqs.~(\ref{eq:resum}) and 
(\ref{eq:NLOresum}) are similar to the ones obtained in 
Refs.~\cite{Chung:2024jfk, Chung:2026mii} for FFs at leading power in $1/N$. 
Because the exponentials $\exp \left[ J^N_{^1S_0^{[1,8]}} \right]$ vanish 
faster than any power of $N$ for large $N$, the inverse Mellin transform of the
resummed expression yields a smooth function of $z$ that vanishes at $z = 1$. 
Note that the resummed expressions are computed at a fixed renormalization
scale for $\alpha_s$. This is because the quantities I compute in
Eqs.~(\ref{eq:resum}) and (\ref{eq:NLOresum}) are NRQCD SDCs,
which must be computed at a fixed matching scale in order to
preserve the universality of the matrix elements. In practice, this scale is
chosen to be of the order of the invariant mass $M$ of the $Q \bar Q$. 

While the expressions for the resummed FFs Eqs.~(\ref{eq:resum}) and
(\ref{eq:NLOresum}) are valid at the scale of order $M$, 
the FFs are usually computed at much higher values of the factorization scale
$\mu_f$ when they are used in calculation of observables. 
The FFs at a higher scale can be obtained by solving the
DGLAP evolution equations. 
At the leading logarithmic level, the effect of DGLAP evolution 
is given by the following factor that multiplies the Mellin-space gluon FF:
\begin{align}
\Delta_{\rm DGLAP}^N = 
\left( \frac{\alpha_s'}{\alpha_s} \right)^{-2 \tilde{P}_{gg}(N)/\beta_0}, 
\end{align}
where $\alpha_s'$ is the strong coupling at the scale $\mu_f$, 
$\tilde{P}_{gg}(N) = 2 C_A \left[ - \psi^{(0)} (N+1) + \psi^{(0)} (1) + 
\frac{1}{N (N-1)} + \frac{1}{(N+1) (N+2)} \right] + \beta_0/2$ is the
tree-level gluon splitting function in Mellin space, 
$\beta_0 = (11 N_c-2 n_f)/3$ is the one-loop beta function of QCD, and $n_f$ is
the number of light quark flavors. 
Here I neglect the effect of quark-gluon mixing, as for the $^1S_0^{[1]}$ 
and $^1S_0^{[8]}$ channels its effect is numerically insignificant and does not
appear in the fixed-order calculation until NNLO in $\alpha_s$. 
At large $N$, this factor behaves like 
\begin{align}
\Delta_{\rm DGLAP}^N \sim N^{-4 C_A \log (\alpha_s/\alpha_s')/\beta_0}, 
\end{align}
Hence, for large $N$ this factor will vanish like some inverse power of $N$. 
Although this vanishes slower than the factor 
$\exp\left[ J^N_{^1S_0^{[1,8]}} \right]$ from threshold resummation, 
this is enough to make the gluon FFs vanish as $z \to 1$ for
$\alpha_s'<\alpha_s$. 
Because the effect of DGLAP evolution has been included for the $^1S_0^{[1]}$
and $^1S_0^{[8]}$ channels in previous phenomenological studies, 
it would be necessary to examine how the effects of threshold resummation and
DGLAP evolution compare and how they combine. I define 
\begin{align}
\label{eq:resumDGLAP}
\tilde{d}_{g \to Q \bar{Q} (^1S_0^{[1,8]})}^{\rm Resum+DGLAP} (N, \mu_f)
&=
\Delta_{\rm DGLAP}^N 
\exp \left[ J^N_{^1S_0^{[1,8]}} \right] 
\tilde{d}_{g \to Q \bar{Q} (^1S_0^{[1,8]})}^{\rm LO} (N), 
\end{align}
which includes both the effects of threshold resummation and DGLAP evolution, 
and 
\begin{align}
\label{eq:NLOresumDGLAP}
\tilde{d}_{g \to Q \bar{Q} (^1S_0^{[1,8]})}^{\rm NLO+ Resum+DGLAP} (N, \mu_f)
&=
\Delta_{\rm DGLAP}^N
\exp \left[ J^N_{^1S_0^{[1,8]}} \right]
\nonumber \\ & \quad \times 
\left( \tilde{d}_{g \to Q \bar{Q} (^1S_0^{[1,8]})}^{\rm NLO} (N)
- J^N_{^1S_0^{[1,8]}}  \tilde{d}_{g \to Q \bar{Q} (^1S_0^{[1,8]})}^{\rm LO} (N)
\right),
\end{align}
which also includes the contribution from the fixed-order NLO correction. 
For comparison, I define the DGLAP-evolved FFs without threshold resummation
\begin{align}
\label{eq:FFDGLAP}
\tilde{d}_{g \to Q \bar{Q} (^1S_0^{[1,8]})}^{\rm LO+DGLAP} (N, \mu_f)
&=
\Delta_{\rm DGLAP}^N
\tilde{d}_{g \to Q \bar{Q} (^1S_0^{[1,8]})}^{\rm LO} (N),
\end{align}
and 
\begin{align}
\label{eq:FFNLODGLAP}
\tilde{d}_{g \to Q \bar{Q} (^1S_0^{[1,8]})}^{\rm NLO+DGLAP} (N, \mu_f)
&=
\Delta_{\rm DGLAP}^N
\tilde{d}_{g \to Q \bar{Q} (^1S_0^{[1,8]})}^{\rm NLO} (N).
\end{align}

\section{Numerical results}
\label{sec:numer}

In this section I present numerical results for the resummed gluon FFs for the 
$^1S_0^{[1]}$ and $^1S_0^{[8]}$ channels. 
I only consider the charmonium case. The bottomonium case is obtained by
changing the heavy quark mass $m$ and $n_f$; the change in $m$ affects the
overall normalization and the scale for the $\alpha_s$, while the change in
$n_f$ affects the fixed-order NLO corrections and the DGLAP evolution kernel. 
Since these changes do not affect the shapes of the resummed FFs appreciably 
I do not show the results for the bottomonium case separately. 

At LO in $\alpha_s$, the gluon FFs for the $^1S_0^{[1]}$ and 
$^1S_0^{[8]}$ channels have the same shapes in $z$ and differ only by the
overall normalization. The FFs at LO in $\alpha_s$ are given by 
\begin{align}
d_{g \to Q \bar{Q} (^1S_0^{[1,8]})}^{\rm LO} (z)
&= {\cal N}^{[1,8]} d_{\rm LO} (z), 
\end{align}
where 
\begin{align}
{\cal N}^{[1]} &= \frac{\alpha_s^2}{4 N_c^2 m^3}, \\
{\cal N}^{[8]} &= \frac{\alpha_s^2 (N_c^2-4)}{4 N_c (N_c^2-1) m^3}, 
\end{align}
and $d_{\rm LO} (z)$ is a dimensionless function in $z$ given by 
\begin{align}
d_{\rm LO} (z) &= (3-2 z ) z +2 (1-z) \log (1-z), 
\end{align}
which satisfies $d_{\rm LO} (z=1) = 1$. 
Since threshold resummation modifies the $z$-shapes of the FFs rather than the
normalization, throughout this section I show in figures the dimensionless
quantities $d_{g \to Q \bar{Q} (^1S_0^{[1,8]})} (z)/{\cal N}^{[1,8]}$ rather
than the FFs themselves. 

\begin{figure}[t]
\includegraphics[width=\columnwidth]{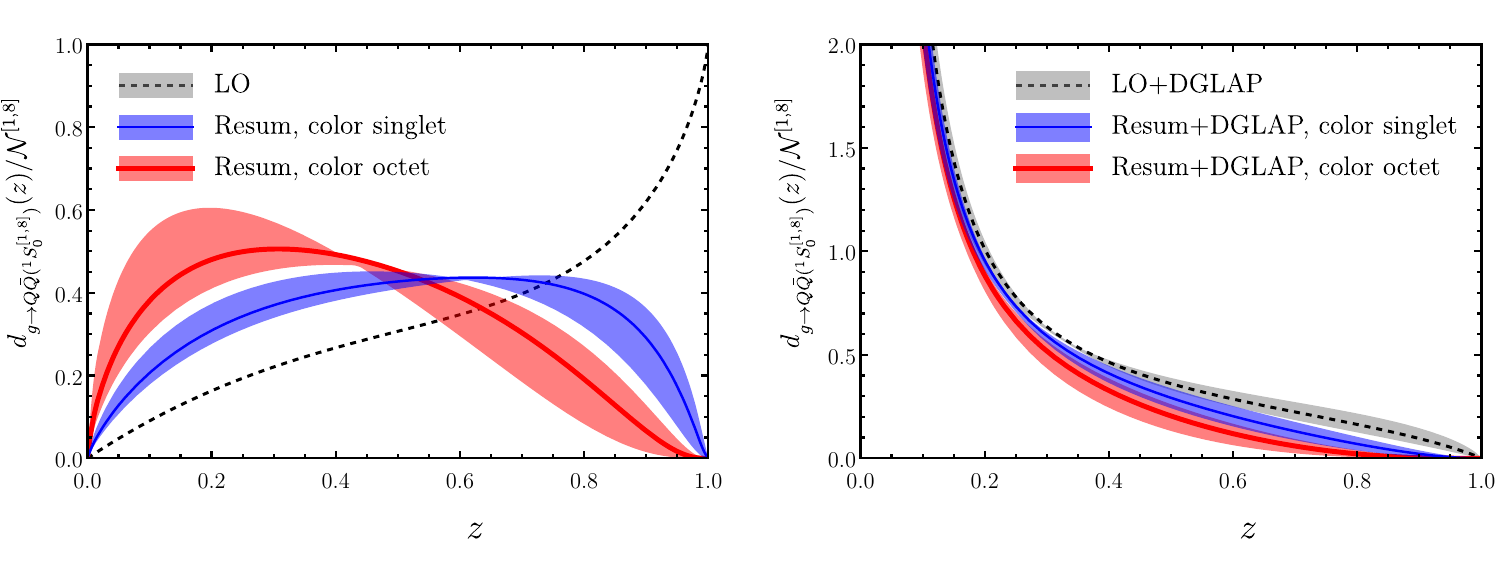}%
\caption{\label{fig:LOResum} 
Left panel: threshold-resummed gluon FFs for $^1S_0^{[1]}$ (blue solid line) 
and $^1S_0^{[8]}$ (thick red solid line) channels, 
 compared to the fixed-order result at LO in $\alpha_s$
(black dotted line). 
Right panel: same as left panel, but with DGLAP evolution to $\mu_f=50$~GeV. 
}
\end{figure}

I set $m = 1.5$~GeV and $n_f =4$. 
The strong coupling $\alpha_s$ is computed at the scale $2 m = 3$~GeV. 
I compute $\alpha_s$ by using {\sf RunDec}~\cite{Herren:2017osy} at four-loop
accuracy to obtain $\alpha_s = 0.254$. 
I consider uncertainties from varying the scale $\mu_r$ for $\alpha_s$ by 
factors $2$ and $1/2$. 
The results for the resummed FFs obtained from Eq.~(\ref{eq:resum}) based on 
the LO calculation are shown in Fig.~\ref{fig:LOResum}. 
Similarly to the results in Refs.~\cite{Chung:2024jfk, Chung:2026mii}, 
the resummed FFs vanish as $z\to 1$,
whereas the LO result has a finite nonzero value at threshold. 
The result for the color-octet channel is more strongly suppressed at
threshold, because the exponent $J^N_{^1S_0^{[8]}}$ is twice as large compared
to the color-singlet case. 
Note that the LO result is not affected by the scale variation, as at LO 
this only changes the overall normalization ${\cal N}^{[1,8]}$, which is
factored out in the results shown in figures. 

Next, I consider the effect of DGLAP evolution. The DGLAP evolution of the
threshold-resummed FF is computed by using Eq.~(\ref{eq:resumDGLAP}).
I choose the scale $\mu_f = 50$~GeV because the cross section becomes dominated
by fragmentation at around this scale. I compute the strong coupling at this
scale by using {\sf RunDec}~\cite{Herren:2017osy} at four-loop
accuracy to obtain $\alpha_s' = 0.130$. 
Similarly to the scale for $\alpha_s$, I also consider the uncertainty from 
varying the scale $\mu_f$ by factors $2$ and $1/2$. 
The effects of variations of the scales $\mu_r$ for $\alpha_s$ and $\mu_f$ are
computed separately, and the uncertainties are added in quadrature. 
The results for 
threshold-resummed FFs evolved to the scale $\mu_f = 50$~GeV are shown in
Fig.~\ref{fig:LOResum}. 
For comparison, I also show the result without threshold resummation
which I compute from Eq.~(\ref{eq:FFDGLAP}).
As previously discussed, DGLAP evolution alone is enough to make the FF vanish
at $z = 1$. However, threshold resummation strongly suppresses the FFs near 
$z = 1$, making the FFs vanish very quickly as $z \to 1$. 
This may be phenomenologically important for differential observables like
the quarkonium momentum distribution in jets, 
especially for testing scenarios where
the cross section is dominated by the spin-singlet channel. 

\begin{figure}[t]
\includegraphics[width=\columnwidth]{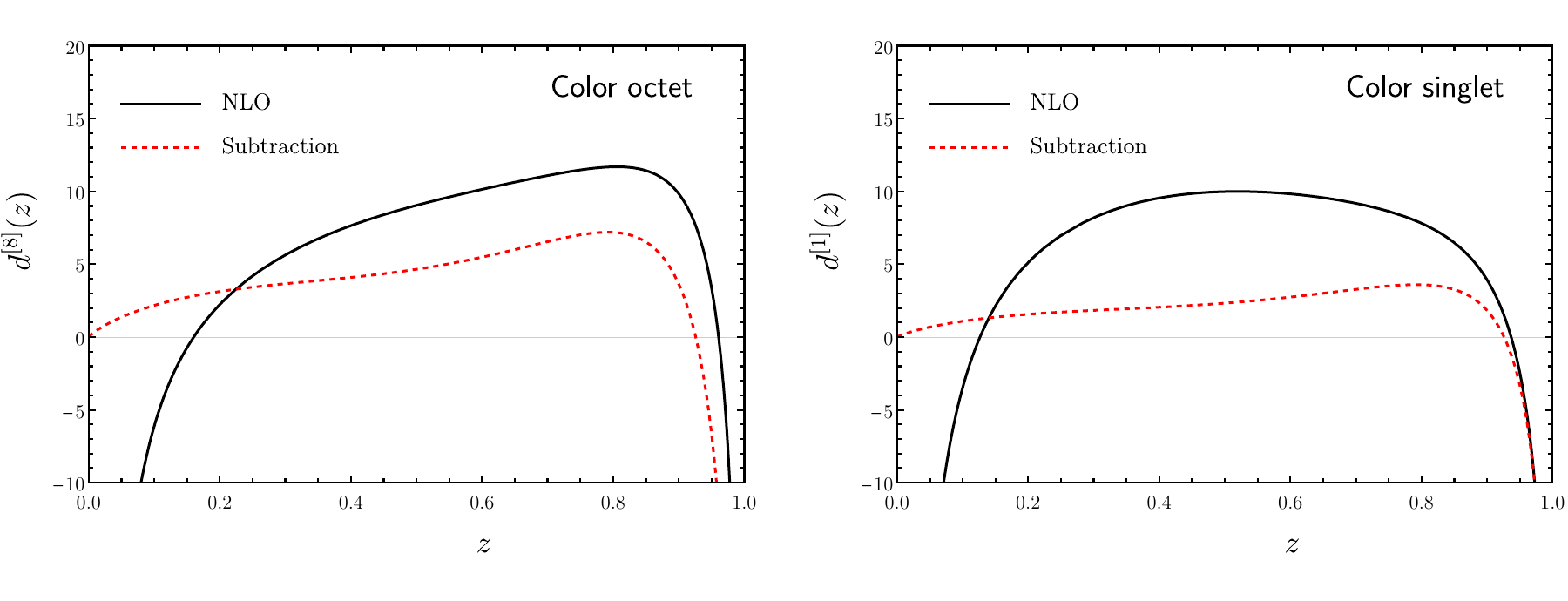}%
\caption{\label{fig:subterm}
Left panel: comparison of the fixed-order NLO correction term $d^{[8]}(z)$ for
the color-octet channel (black solid line) with the subtraction term given by 
Eq.~(\ref{eq:subterm}) times $C_A$ (red dotted line). 
Right panel: comparison of the fixed-order NLO correction term $d^{[1]}(z)$ for
the color-singlet channel (black solid line) with the subtraction term given by
Eq.~(\ref{eq:subterm}) times $C_A/2$ (red dotted line).
}
\end{figure}

Next, I consider the effect of inclusion of the fixed-order NLO correction. 
If I set $\mu_f = M$ and also compute $\alpha_s$ at the same scale, the
fixed-order FFs can be written as
\begin{align}
d_{g \to Q \bar{Q} (^1S_0^{[1,8]})}^{\rm NLO} (z)
&= {\cal N}^{[1,8]} \left[ d_{\rm LO} (z)
+ \frac{\alpha_s}{\pi} d^{[1,8]} (z) \right], 
\end{align}
where the functions $d^{[1]} (z)$ and $d^{[8]} (z)$ have been computed semi
analytically in Ref.~\cite{Zhang:2018mlo}. 
These functions contain the double logarithmic terms 
$- \frac{1}{2} N_c \log^2(1-z)$ and $- N_c \log^2(1-z)$ for 
color singlet and color octet, respectively. This can be compared with the
subtraction term in Eq.~(\ref{eq:NLOresum}) to see how much of the fixed-order
NLO correction comes from the threshold double logarithm. 
The subtraction term can actually be computed analytically in $z$ space. 
For color octet, the subtraction term is given by 
${\cal N}^{[8]} \frac{\alpha_s C_A}{\pi} 
\left[ \frac{-2 \log(1-z)}{1-z} \right]_+ \otimes d_{\rm LO}(z)$, with
\begin{align}
\label{eq:subterm}
\left[ \frac{-2 \log(1-z)}{1-z} \right]_+ \otimes d_{\rm LO}(z)
&= 
z (2 z-3) \left[ \log ^2(1-z) +2 \, \text{Li}_2(z)-\frac{\pi^2}{3} \right]
\nonumber \\ & \quad
+4 (1-z) \bigg[ 
- \log ^3(1-z)
+ \log ^2(1-z)
\nonumber \\ & \hspace{15ex}
+\left( \frac{\pi^2}{3} -2 -\log z + \frac{3}{2} z
\right) \log (1-z)
\nonumber \\ & \hspace{15ex}
- \text{Li}_3(1-z) -2 \, \text{Li}_3(z)
- \frac{3}{2} z \log z
\nonumber \\ & \hspace{15ex}
-
\left( \log(1-z)-\log z \right) \text{Li}_2(z)
+ \zeta (3) - \frac{z}{2}
\bigg]
\nonumber \\ & \quad
-4 \, \text{Li}_2(z) 
-4 z \log z
+\frac{2 \pi ^2}{3} z,
\end{align}
and for color singlet, the subtraction term is given by 
$\frac{1}{2} \frac{\alpha_s C_A}{\pi} {\cal N}^{[1]}
\left[ \frac{-2 \log(1-z)}{1-z} \right]_+ \otimes d_{\rm LO}(z)$.
After factoring out the common factors, the function $d^{[8]}(z)$ should be
compared with Eq.~(\ref{eq:subterm}) times $C_A$, and $d^{[1]}(z)$ should be 
compared with Eq.~(\ref{eq:subterm}) times $C_A/2$. 
I show the comparison in Fig.~\ref{fig:subterm}.
For the color-singlet case, the subtraction term almost completely reproduces
the NLO correction term $d^{[1]}(z)$ near threshold, although 
the corrections are still significant for intermediate values of $z$. 
For the color-octet case, the agreement between $d^{[8]}(z)$ and the
subtraction term is not as good as the color-singlet case; even though the
subtraction term reproduces the leading double logarithmic $\log^2 (1-z)$ term 
in $d^{[8]}(z)$ exactly, the contributions from the single logarithms and terms
finite at $z=1$ may still be numerically significant. 
Interestingly, for intermediate values of $z$ the difference between
$d^{[8]}(z)$ and the subtraction term is smaller compared to the color-singlet
case.

\begin{figure}[t]
\includegraphics[width=\columnwidth]{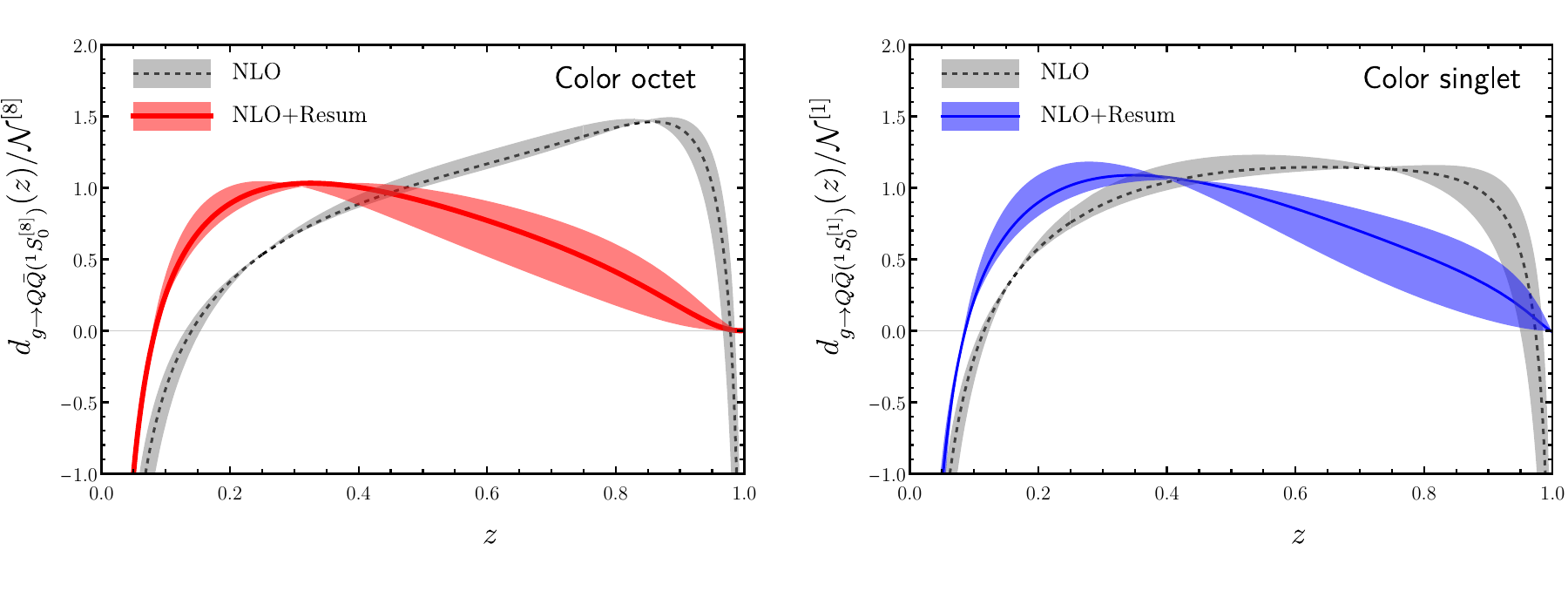}%
\caption{\label{fig:NLOresum}
Left panel: 
threshold-resummed gluon FFs for the $^1S_0^{[8]}$ channel 
(red solid line) compared to the fixed-order result at NLO in $\alpha_s$
(black dotted line).
Right panel: 
threshold-resummed gluon FFs for the $^1S_0^{[1]}$ channel 
(blue solid line) compared to the fixed-order result at NLO in $\alpha_s$
(black dotted line).
The NLO results shown here include the LO contributions. 
}
\end{figure}

Now I show the results for the resummed FFs including the fixed-order NLO
corrections, computed by using Eq.~(\ref{eq:NLOresum}). 
The results are shown in Fig.~\ref{fig:NLOresum}. 
The uncertainties come from scale variations; the uncertainties in the
fixed-order NLO results appear smaller than what are shown in
Ref.~\cite{Zhang:2018mlo}, because the overall normalization ${\cal N}^{[1,8]}$
is factored out in these results. 
Similarly to the LO case, threshold resummation makes the FFs smoothly vanish
as $z \to 1$, whereas the fixed-order NLO results change sign and diverge
double logarithmically at threshold. 

\begin{figure}[t]
\includegraphics[width=\columnwidth]{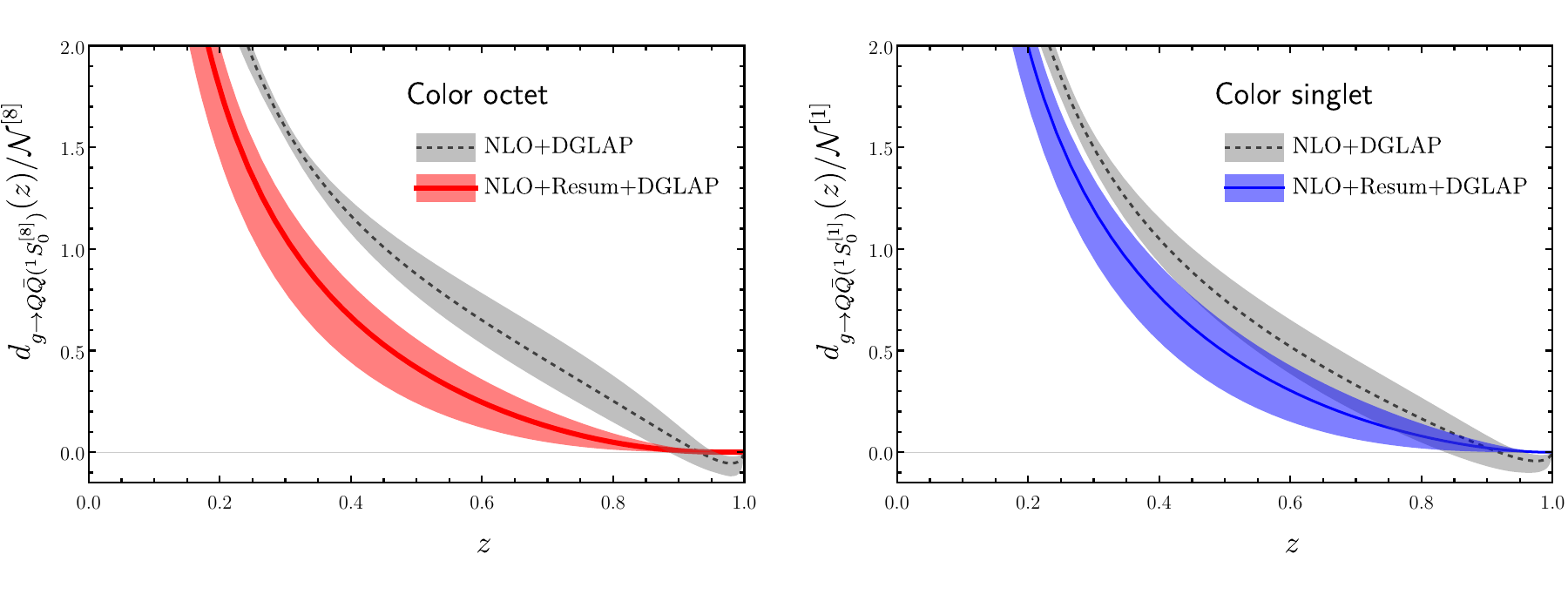}%
\caption{\label{fig:NLOresumDGLAP}
Left panel:
threshold-resummed gluon FFs with DGLAP evolution for the $^1S_0^{[8]}$ channel
(red solid line) compared to the DGLAP-evolved fixed-order result at NLO in 
$\alpha_s$ (black dotted line).
Right panel:
threshold-resummed gluon FFs with DGLAP evolution for the $^1S_0^{[1]}$ channel
(blue solid line) compared to the DGLAP-evolved fixed-order result at NLO in 
$\alpha_s$ (black dotted line).
The NLO results shown here include the LO contributions. 
}
\end{figure}

Next I show the results with DGLAP evolution to $\mu_f = 50$~GeV. 
The results are shown in Fig.~\ref{fig:NLOresumDGLAP}. 
Again, the uncertainties come from variations of $\mu_r$ and $\mu_f$, added in
quadrature. 
Even though the DGLAP evolution makes the fixed-order FFs vanish at $z=1$, 
the fixed-order results change sign near threshold 
due to the negative double logarithmic correction term at NLO. 
On the other hand, in the threshold-resummed results, 
the double logarithmic correction terms in
the fixed-order NLO correction are subtracted and included in the resummed 
exponents $\exp\left[ J^N_{^1S_0^{[1,8]}} \right]$. As a result, the resummed 
FFs remain positive definite near threshold even after including the
fixed-order NLO corrections. 

In the results shown in Fig.~\ref{fig:NLOresumDGLAP}, 
the DGLAP evolution is computed at the leading-logarithmic level, 
even though the NLO fixed-order
corrections are taken into account. While it would be desirable to consider
DGLAP evolution also at next-to-leading logarithmic accuracy to match the 
NLO fixed-order corrections, 
a consistent treatment of threshold resummation and DGLAP evolution requires a
joint resummation after matching the single logarithmic contributions. 
This is currently not possible since this is the first work on the threshold
double logarithms at leading double logarithmic level, neither the NLO double
logarithm nor the single logarithms are known. 
Since DGLAP resummation is typically considered at leading logarithmic level 
in quarkonium production phenomenology, the results shown here can still be
useful in considering the effect of threshold resummation in phenomenology.

\begin{figure}[t]
\includegraphics[width=\columnwidth]{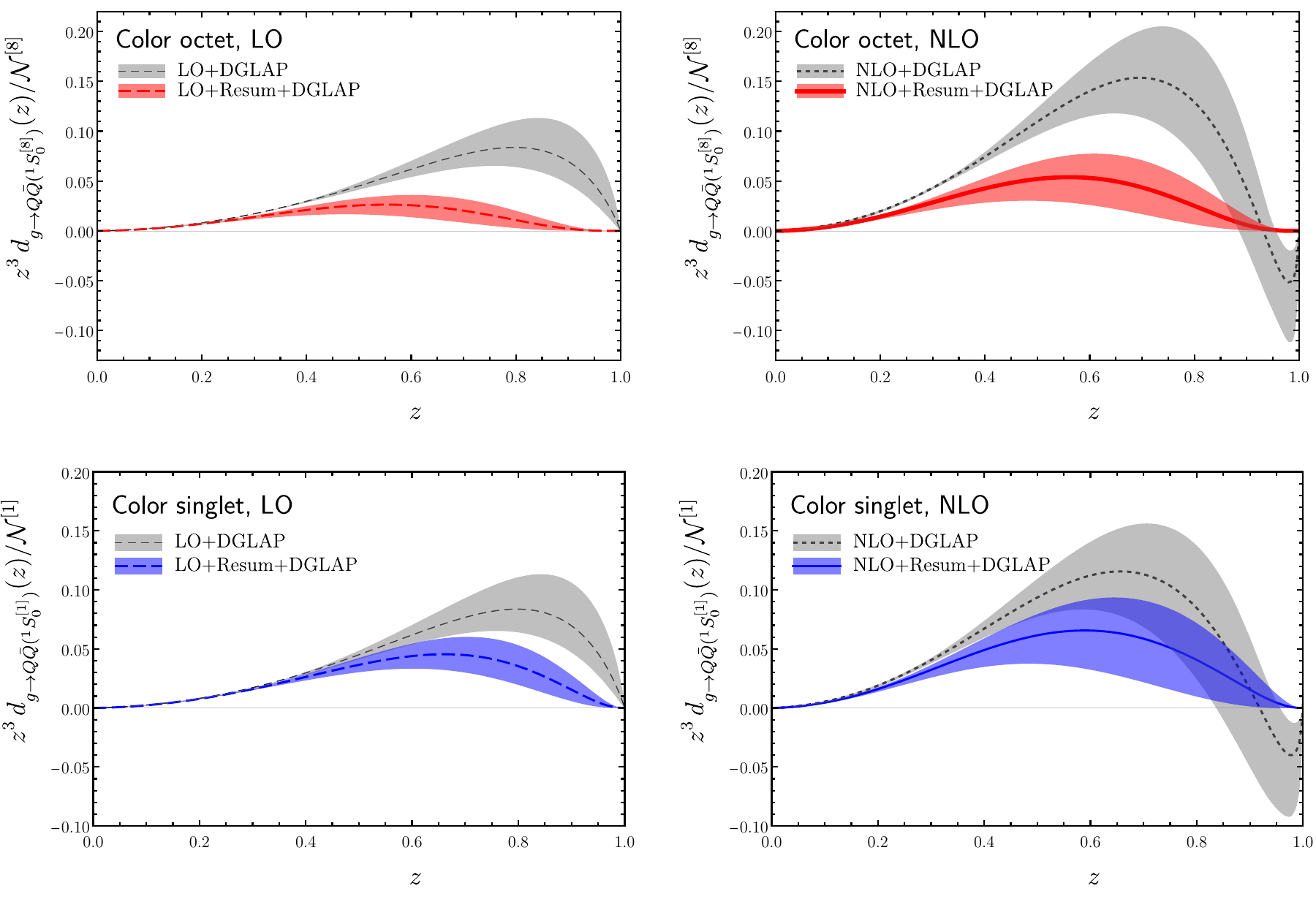}%
\caption{\label{fig:resumDGLAPcompare}
DGLAP-evolved threshold-resummed gluon FFs times $z^3$ 
with fixed-order NLO corrections (solid lines) and without NLO corrections
(dashed lines), compared to DGLAP-evolved gluon FFs times $z^3$ 
without threshold resummation at LO (thin dashed lines) and 
NLO (dotted lines) accuracies. 
The color-octet case is shown on the upper panels, 
and the color-singlet case is shown on the lower panels. 
The NLO results shown in the right panels include the LO contributions. 
}
\end{figure}

Note that the DGLAP-evolved FFs diverge positively as $z \to 0$, 
which comes from the $2 C_A/z$ term in the gluon splitting function. 
In practice, the small-$z$ region does not contribute to physical cross
sections because the momentum of the fragmenting gluon goes to infinity as 
$z \to 0$, and also because the transverse-momentum dependent gluon cross
section rises approximately like $z^3$. Hence, when trying to see how the
$z$-shapes of the FFs affect the cross sections, it can be more useful to
examine $z^3$ times the FFs rather than the FFs themselves. 
In Fig.~\ref{fig:resumDGLAPcompare} I show the threshold-resummed, 
DGLAP-evolved FFs at the scale
$\mu_f = 50$~GeV multiplied by $z^3$, with and without the fixed-order NLO
corrections. For comparison, I also show the DGLAP-evolved FFs times $z^3$
without threshold resummation at both LO and NLO accuracies. 
The uncertainties again come from variations of $\mu_r$ and $\mu_f$, added in
quadrature. 
It is clear that even after DGLAP evolution, 
without threshold resummation the fixed-order NLO
corrections have significant effects to the $z$-shapes of the FFs especially in
the threshold region, where the double logarithmic corrections make the FFs 
change sign as $z \to 1$. 
This unphysical behavior goes away when the threshold double
logarithms are resummed, and the fixed-order NLO corrections do not appreciably
change the $z$ shapes of the FFs. While the NLO corrections can still be
numerically significant in the threshold-resummed case, the bulk of the effect
is to change the overall normalization of the FFs, especially for the
color-octet case. 
If the nonperturbative matrix elements are determined by comparing to cross
section data, as is usually the case, 
this change in normalization is unlikely to have a significant phenomenological
impact. 
However, if the matrix elements are constrained from other methods, such as
potential-model calculations or heavy-quark spin symmetry, the NLO corrections
may bring in changes in the sizes of the $^1S_0^{[1,8]}$ channel contributions 
by almost a factor of 2. 

Even though the results for the FFs shown in this section are computed at a
fixed scale for $\alpha_s$, in general the smallest typical scale of the 
exchanged gluon
in the threshold region $z \to 1$ is the soft scale $\mu_s=M(1-z)$. 
In the charmonium case with $M =3$~GeV, this scale is in the nonperturbative
region $\mu_s \lesssim 500$~MeV for $z \gtrsim 0.8$. In such case,
nonperturbative effects may dominate the $z$-shape of the FF. 
While threshold resummation suggests that the FFs will vanish smoothly as 
$z \to 1$, this does not automatically exclude the possibility for features to
appear in the threshold region due to nonperturbative effects. 
For production of a physical quarkonium state, nonperturbative effects in the 
$z$-shape of the FFs can be described in terms of the shape
functions~\cite{Beneke:1997qw, Fleming:2003gt}. 
While currently it is not known how to compute the shape functions for
quarkonium production processes, they may be probed experimentally from
differential observables such as quarkonium production in jet.

\section{Summary and Discussion}
\label{sec:concl}

In this work I resummed the threshold double logarithms that appear in
radiative corrections to the gluon FFs for production of a quark and antiquark
pair in the $S$-wave spin-singlet state, for both the color-octet and
color-singlet cases. 
This is based on the analysis of soft gluon emissions that can be organized 
into soft functions built from Wilson lines. 
By using the fact the threshold double logarithms arise from planar diagrams in
loop corrections to the soft functions, the resummation is achieved by
exponentiation in Mellin space. 
This is the first result for threshold resummation for gluon FFs for quarkonium
production for channels that are suppressed by $1/N$ in Mellin space. 

Similarly to the case of gluon FFs for the $^3S_1^{[8]}$ and $^3P_J^{[1,8]}$
channels, threshold resummation has a significant impact on the $z$-shapes of
the gluon FFs for the $^1S_0^{[1,8]}$ channels, especially when the
next-to-leading order corrections in $\alpha_s$ are included. The double 
logarithmic correction terms in the fixed-order calculation make the FFs change
sign near threshold, producing unphysical negative contributions as $z \to 1$. 
In threshold-resummed FFs, the double logarithmic correction terms are instead
included in the resummed exponent, which vanishes rapidly for large $N$ in
Mellin space. The resulting $z$-space FFs vanish smoothly as $z \to 1$. 
Even when compared to the leading-order FF, threshold resummation makes the
$z$-shapes of the FFs much softer; this can be phenomenologically important
for differential observables if the production mechanism is dominated
by the spin-singlet state. On the other hand, threshold resummation
diminishes the $z$ dependence of the next-to-leading order corrections. 
As the bulk of the effect of the corrections is to
change the overall normalizations of the threshold-resummed FFs, 
if the corresponding matrix element is obtained phenomenologically by 
comparing to cross section data, the NLO corrections to threshold-resummed FFs 
are unlikely to have a significant phenomenological impact. 
However, if the matrix elements are further constrained from other methods, 
the contribution from the $^1S_0^{[1,8]}$ channels may be enhanced by almost a
factor of 2.

Because the gluon FFs for the $^1S_0^{[1,8]}$ channels are suppressed by $1/N$
in Mellin space, DGLAP evolution alone is sufficient to make the FFs vanish at
threshold. 
However, even with DGLAP evolution, threshold resummation considerably softens
the $z$ shapes of the FFs. Hence, threshold resummation may be
phenomenologically important for studies of differential observables such as 
the quarkonium momentum distribution in jets, if the production mechanism is
dominated by the spin-singlet channel. However, if the spin-singlet channel is
only a part of the production mechanism, which is the case for many scenarios
for $J/\psi$ production, the effect of threshold resummation 
for this channel may not be so significant in the sum over all channels, 
because the $^3S_1^{[8]}$ and $^3P_J^{[1,8]}$ channels tend to produce much
harder distributions in $z$ even after threshold resummation.

The method developed in this work to resum threshold double logarithms for
gluon FFs that are suppressed by powers of $1/N$ may be applied to other
channels with similar large $N$ behaviors. For example, in the case of
$^3S_1^{[8]}$ and $^3P_J^{[1,8]}$ channels, longitudinally polarized
contributions begin to appear from order $\alpha_s^2$, which also vanish like 
$1/N$ for large $N$~\cite{Braaten:2000pc, Ma:2015yka}. Threshold resummation
for these channels may be important
for precision calculation of quarkonium polarization. 
Moreover, the method may also be extended to cases where multiple 
collinear gluons are necessary to produce a quark and antiquark pair in a
specific color and angular momentum state. For example, gluon fragmentation
for the $^3S_1^{[1]}$ channel occurs through emission of two gluons, which is
necessary in producing a color-singlet state with spin one from fragmentation
of a gluon~\cite{Braaten:1993rw, Braaten:1995cj, Braaten:1996rp, 
Bodwin:2003wh, Bodwin:2012xc, Zhang:2017xoj}. 
Although the gluon FF for this channel at the lowest power in the
nonrelativistic expansion already vanishes as $z \to 1$, 
the relativistic correction to this channel is known to be finite
and nonzero at $z=1$~\cite{Bodwin:2003wh}. 
Since this channel involves one more lightlike direction
than the spin-singlet case due to the extra collinear gluon in the
final state, it is possible that the exponent for threshold resummation may
even be larger compared to the $^1S_0^{[1]}$ channel. 
If this is the case, threshold resummation may substantially soften the
$z$ shape of the relativistic correction to the gluon FF for the $^3S_1^{[1]}$
channel, which may help bring the nonrelativistic expansion of color-singlet
spin-triplet quarkonium production under control.

\begin{acknowledgments}
This work was supported by the National Research Foundation of Korea (NRF)
grant funded by the Korean government (MSIT) in 2026 (Grant No.
RS-2026-25497906, Project Title: Hadron Production at Next-Generation
Colliders).
\end{acknowledgments}

\bibliography{FF1S0_paper.bib}

\end{document}